\documentclass[twocolumn]{revtex4}
\usepackage{epsfig}
\usepackage{graphicx}
\usepackage{amsmath}
\usepackage{mathrsfs}
\usepackage{dcolumn}
\usepackage{amsfonts}
\usepackage{ulem}
\newcommand{\Tr}{\text{Tr}}
\newcommand{\ket}[1]{|#1\rangle}
\newcommand{\bra}[1]{\langle#1|}

\begin{document}
 \title{Non-Markovianity and memory effects in quantum open systems}
 \author{S. C. Hou$^{1,2}$, S. L. Liang$^{1}$, and X. X. Yi$^{1}$}

\affiliation{$^{1}$ Center for Quantum Sciences and School of
Physics, Northeast Normal University, Changchun 130024, China\\
$^{2}$ School of Physics and Optoelectronic Technology, Dalian
University of Technology, Dalian 116024, China}

\begin{abstract}
Although a number of measures for quantum non-Markovianity
have  been proposed recently,  it is still an open question whether
these measures directly  characterize the memory effect of the
environment, i.e., the dependence of a quantum state on its past in
a time evolution. In this paper, we present  a criterion and propose
a measure for  non-Markovianity with  clear physical
interpretations of the  memory effect. The non-Markovianity
is defined  by the inequality $T(t_2,t_0)\neq T(t_2,t_1)T(t_1,t_0)$ in
terms of  memoryless dynamical map $T$ introduced  in this paper.
This definition  is conceptually distinct from that based on
divisibility used by Rivas \textit{et al} (Phys. Rev. Lett
 105, 050403 (2010)), whose violation is manifested by  non-complete
positivity of the dynamical map. We demonstrate via a
typical quantum process that without Markovian approximation,
nonzero memory effects (non-Markovianity) always exist even if the
non-Markovianity is zero by the other non-Marovianity measures.
\end{abstract}
\date{\today}
\maketitle

\section{Introduction}
The Markov approximation is widely used to study open  quantum
systems, leading to a  reduced dynamics where the future state  of
the system depends only on its present state. Mathematically, a
quantum Markovian process can be described  by a master equation in
the Lindblad form, or equivalently by completely positive divisible
maps\cite{Breuerbook,Rivas}. Strong-coupling, finite-size
environments or small time scales might lead to the failure of the
Markov approximation. Memory effects then become important, and the
dynamics in this case is said to be non-Markovian. It was found that
the non-Markovianity is usually associated with the occurrence of
revivals, non-exponential relaxation, or negative decay rates in the
dynamics. In recent years, the features of the non-Markovian quantum
process attracted attention in both theoretical and experimental
studies \cite{Chruscinski,Jing,Fanchini,Xu,Madsen,Liu,Zhang,
Chin,Huelga,Deffner,Kennes,Tahara,Cerrillo,Shabani}. A number of
quantitative measures have been proposed to quantify
non-Markovianity, such as the quantum channels \cite{Wolf}, the
non-monotonic behaviors of distinguishability
\cite{Breuer,Laine,Vasile,Liuj}, entanglement \cite{Rivas2}, Fisher
information\cite{Lu}, correlation \cite{Luo}, the volume of states
\cite{Lorenzo}, capacity \cite{Bylicka}, the breakdown  of
divisibility \cite{Rivas2,Hou}, the negative fidelity difference
\cite{Rajagopal}, the nonzero quantum discord \cite{Alipour}, the
negative decay rates \cite{Hall}, and the notion of non-Markovian
degree \cite{Chruscinski2014}.

Among them, the measure defined by Breuer, Laine and Piilo
\cite{Breuer} is closely related to the memory effect:
Non-Markovianity in this measure manifests itself as a reverse flow
of information from the environment  to the system. This back-flow
of information  might be a sufficient condition for the
memory effects. It is worth addressing that the essence of the
non-Markovianity is whether  {\it the future state  depends on its
past state or not}. To that extent, there are no straightforward
witnesses  for non-Markovianity to date, although a great deal of
effort has been made to understand the non-Markovianity. The
reason that previous  measures
\cite{Wolf,Breuer,Laine,Vasile,Liuj,Rivas2,Lu,Luo,Lorenzo,
Bylicka,Hou,Rajagopal,Alipour,Hall,Chruscinski2014} might  not
capture directly the memory effect is the physics behind the
definitions. The measure in Ref.\cite{Wolf} investigates whether a
quantum channel is consistent with a Markovian evolution. The
measure proposed by Rivas, Huelga, and Plenio uses divisibility as a
measure of non-Markovianity which characterizes the noncomplete
positivity of the dynamical map \cite{Rivas2}. The noncomplete
positivity is improved   in \cite{Hou}, but it still could not
capture the memory effects. The other non-Markovianity measures in
Refs. \cite{Breuer,Laine,Vasile,Liuj,Rivas2,Lu,Luo,Lorenzo,
Bylicka,Rajagopal,Alipour,Hall,Chruscinski2014} are based on
 features  different  from those  mentioned before, however these features  are
intrinsically related to the non-complete positivity of dynamical
maps in quantum evolutions. Therefore, the  measures so far do not
directly reflect how the future state  of a quantum system depends
on its past.   This stimulates us to consider the following
questions:  How do we directly quantify this memory effect? Is
non-exponential but monotonic relaxation a Markovian process or a
non-Markovian one? What is the essential difference between a
time-dependent Markovian equation and a non-Markovian time-local
equation?  In this paper, we will try to study these questions by
quantifying the non-Markovianity directly based on the memory
effect.

Generally speaking,  the conventional  dynamical map
$\varepsilon(t_2,t_1)$ is ambiguous unless the initial time of the
evolution is fixed. We define $T(t_2,t_1)$ (for details, see
below) as the dynamical map transferring state from $\rho(t_1)$ to
$\rho(t_2)$, where $t_1$ (arbitrary)  is the initial time of the
evolution. The initial time means the state of the whole system is
in a product state at this point in time. We will show the
difference between $T(t_2,t_1)$ and $\varepsilon(t_2,t_1)$ in latter
discussions. By this definition, $T$ is always completely positive
(CP) and trace preserving (TP) for arbitrary $\rho(t_1)$ in quantum
processes. The Markovian divisibility condition can be expressed in
terms of dynamical map $T$ as $T(t_2,t_0)= T(t_2,t_1)T(t_1,t_0)$
because the dynamical map in a Markovian process does not
depend on the initial time of evolution. In the non-Markovian
regime, the divisible condition in terms of $T$ is conceptually
distinct from the conventional one since $T(t_2,t_1)$ is no longer a
dynamical map connecting $T(t_1,t_0)$ and $T(t_2,t_0)$. Instead,
$T(t_2,t_1)$ and $T(t_2,t_0)$ correspond to two evolutions with
different initial times in a quantum process. Our main
result  is that the non-Markovianity in terms of $T$ is manifested
by   inequality $T(t_2,t_0)\neq T(t_2,t_1)T(t_1,t_0)$ which has a
physical explanation that the evolution depends on its history.
This is distinct from the non-Markovianity measure defined
via completely positive divisibility discussed in
Refs.\cite{Rivas2,Hou,Chruscinski2014} for the dynamical map.

With those notations, the non-Markovianity is then defined as the
maximum distance between
 $T(t_2,t_0)$ and $T(t_2,t_1)T(t_1,t_0)$ over $t_1$ and $t_2$.
It reflects the degree of memory effects in a quantum process. We
calculate the non-Markovianity of a typical quantum process where a
qubit decays into vacuum with different environmental
spectra. The results show that a non-zero memory effect
(non-Markovianity) always exists in this process, even if it is
Markovian by the previous measures. Our measure applies to a quantum
process that can describe evolutions starting from an arbitrary
time $t_I$, rather than an evolution with a fixed initial time.

The paper is organized  as follows. In Sec.\rm{II},  we review the
dynamics of open quantum systems and the concept of universal
dynamical map. In Sec.\rm{III}, we define two types of dynamical
maps, $\Lambda$ and $T$, with different physical meanings. The
non-Markovian criterion and measure is introduced in Sec.\rm{IV} in
terms of $T$. In Sec.\rm{V}, we calculated and discuss the
non-Markovianity of an exactly solvable model. Finally, we summarize
 in Sec.\rm{VI}.

\section{dynamics of open quantum systems }
A natural way to model an open quantum system is to  regard it as
arising from an interaction between the system and the environment
(denoted by $S$ and $E$), which together form a closed quantum
system. The reduced density matrix for the system is obtained by
tracing out the environmental degrees of freedom, i.e.,
$\rho_S=\Tr_E(\rho_{SE})$. The total Hamiltonian for the system and
the environment can be written as,
\begin{eqnarray}
H=H_S+H_E+H_{SE},
\end{eqnarray}
where $H_S$, $H_E$ and $H_{SE}$ represents the  Hamiltonian of the
system, the environment and the coupling, respectively. Consider the
total density matrix at $t_1$,  $\rho_{SE}(t_1)$, which undergoes a
unitary evolution. The system density matrix at $t_2$ ($t_2\geq
t_1$) is given by
\begin{eqnarray}
\rho_S(t_2)=\Tr_{E}[U(t_2,t_1)\rho_{SE}(t_1) U(t_2,t_1)^{\dag}],
\label{Eqn:gmap}
\end{eqnarray}
where $U(t_2,t_1)=e^{-\frac{i}{\hbar}H(t_2-t_1)}$. In  a general
case where the total Hamiltonian is time-dependent, $U(t_2,t_1)$ can
be expressed as
$U(t_2,t_1)=\mathcal{T}e^{-\frac{i}{\hbar}\int_{t_1}^{t_2}H(\tau)d\tau}$
with $\mathcal{T}$ the chronological operator. Eq.(\ref{Eqn:gmap})
can be understood as a map $\varepsilon(t_2,t_1)$ connecting
$\rho_S(t_1)$ and $\rho_S(t_2)$, namely,
\begin{eqnarray}
\rho_S(t_2)=\varepsilon(t_2,t_1)\rho_S(t_1).
\label{Eqn:map}
\end{eqnarray}
In general, such a map depends not only on the total  evolution
operator $U(t_2,t_1)$ and properties of the environment, but also on
the system state  because $\rho_{SE}(t_1)$ may contain correlations
between the system and the environment \cite{Rivas}. Given a fixed
$\rho_E(t_1)$ of the environment and the correlations between $S$ and $E$, not all
$\rho_S(t_1)$ are allowed due to the positivity requirement of
$\rho_{SE}(t_1)$. For example, if the correlation is non-zero,
$\rho_S(t_1)$ can not be a pure state \cite{Rivas}. Moreover, it is
well known that $\varepsilon(t_2,t_1)$ may not be CP.

A dynamical map which is independent of the state  it acts upon is
called a universal dynamical map (UDM).  It describes a plausible
evolution for any states\cite{Rivas}. Being TP and CP, a UDM can be
expressed in the Kraus representation,
\begin{eqnarray}
\rho_S(t_2)&=& \varepsilon(t_2,t_1)\rho_S(t_1)\nonumber\\
           &=& \sum_{\alpha} K_\alpha(t_2,t_1)
           \rho_{S}(t_1)K^{\dag}_\alpha(t_2,t_1),
           \label{Eqn:Kraus}
\end{eqnarray}
where
$\sum_{\alpha}K^{\dag}_\alpha(t_2,t_1)K_\alpha(t_2,t_1)=\mathbb{I}$.
It turns out that a dynamical map is a UDM if and only  if it is
induced from the initial condition
$\rho_{SE}(t_1)=\rho_S(t_1)\otimes\rho_E(t_1),$ where $\rho_E(t_1)$
is fixed (independent of the system state) for arbitrary
$\rho_S(t_1)$ \cite{Rivas}.

Notice that in a Markovian or non-Markovian quantum process,  an
evolution could start not only at $t=0$, but also at the other
times, say, $t=t_I$ ($t_I\geq0$) where we refer to $t_I$ as
the initial time of the evolution. Starting at $t_I$ implies that the
total state of the system and the environment satisfies  conditions at
$t_I$. In this paper, we focus on quantum processes where the
evolutions start with an initial product state of the system and
the environment, i.e., $\rho_{SE}(t_I)=\rho_{S}(t_I)\otimes\rho_E(t_I)$
where $\rho_E(t_I)$ is fixed and $\rho_{S}(t_I)$ is arbitrary. In
other words, we study models describing a system in arbitrary states
interacting with the environment in a fixed initial state.
This consideration  is reasonable as it means that  the
system and the environment are  independent at the beginning $t_I$ and
then start their evolution. Typically, $\rho_E(t_I)=\rho_E$ is
stationary under a time-independent environment Hamiltonian $H_E$,
i.e.,  $\rho_{SE}(t_I)=\rho_S(t_I)\otimes\rho_E$ for any $t_I$. For
example, the equilibrium state or the vacuum state of the environment is
often chosen as an initial environment state to study the reduced
system dynamics. For a general (nonstationary) initial
environment state, it can be given by
\begin{eqnarray}
\rho_E(t_I)&=& \mathcal{T}e^{-\frac{i}{\hbar}\int_{0}^{t_I}H_E(\tau)d\tau}\rho_E(0).
\label{Eqn:ini_env}
\end{eqnarray}
where we have assumed that before an evolution starts, initial state
of the environment undergoes free evolution driven  by $H_E$
[$H_E(t)$].

With the factorized initial state, the dynamical maps starting from
any time $t_I$ are UDMs characterized by the fixed initial
environment state and the global unitary evolution (total
Hamiltonian), rather than the initial system state. Consequently,
the quantum processes are supposed to have fixed non-Markovianity as
a function of $\rho_E(t_I)$ and $H(t)$, regardless of the system
states. On the other hand, the quantum process is universal because
it can describe the evolution of any initial system state. In the case
of $\rho_E(t_I)=\rho_E$ and $H(t)=H$, which is the usual assumption
used to investigate the dynamics of an open quantum system, we expect a
time-independent value of non-Markovianity of the quantum process.
Otherwise, the non-Markovianity may depend on the time interval that
we are interested in since the quantum process is not time-
homogenous.

\section{Dynamical maps with and without memory}

The reduced dynamics of an open quantum system can be  represented
by dynamical maps. In particular, quantum non-Markovian behaviors are
often discussed with the help of the dynamical map
$\varepsilon(t_2,t_1)$ that transforms the state from  $\rho_S(t_1)$ to
$\rho_S(t_2).$  However, in general the physical meaning and
properties of $\varepsilon(t_2,t_1)$ are not clear unless we know
the total density matrix $\rho_{SE}(t_1)$, or, from another point of
view, the initial time of the evolution. For instance, let $t_0\leq
t_1\leq t_2$; when the evolution starts at $t_1$, which means
$\rho_{SE}(t_1)=\rho_S(t_1)\otimes\rho_E$, the map
$\varepsilon(t_2,t_1)$ is a UDM. In contrast, when the evolution
starts at a fixed time $t_0$, the total density matrix at $t_1$ is
$\rho_{SE}(t_1)=U(t_1,t_0)\rho_{S}(t_0)\otimes\rho_E
U(t_1,t_0)^{\dag}$, and then $\varepsilon(t_2,t_1)$ can be understood as
an intermediate map that is in general not a UDM. Obviously,
$\varepsilon(t_2,t_1)$ has different meanings and properties for
different initial times of evolution. This is crucial for
understanding non-Markovian characters since it tells us that the
evolution of the system after $t_1$ [determined by
$\varepsilon(t_2,t_1)$] is related to its history (from $t_0$ to
$t_1$). To clarify this point, we introduce two types of dynamical
maps denoted by $\Lambda$ and $T$, respectively. We will describe
the two dynamical maps in the following.

\subsection{Type \rm{I}: dynamical map with memory}
$\Lambda(t_2,t_1)$ is defined as the quantum dynamical  map
transferring the state from $\rho_S(t_1)$ to $\rho_S(t_2)$ when the
evolution starts at a prescribed time $t_0$ ($t_0\leq t_1\leq t_2$),
i.e., $t_I=t_0$. When $t_1>t_0$, $\Lambda(t_2,t_1)$ can be
understood as an  continuous map. The physical meaning of
$\Lambda(t_2,t_1)$ can be expressed as
\begin{eqnarray}
 \rho_S(t_2)&=&\Lambda(t_2,t_1)\rho_S(t_1)\nonumber
\\&=&\Tr_E[U(t_2,t_1)\rho_{SE}(t_1)U(t_2,t_1)^{\dag}],
\label{Eqn:lambda}
\end{eqnarray}
where $\rho_{SE}(t_1)=U(t_1,t_0)\rho_{S}(t_0)
\otimes\rho_{E}(t_0)U(t_1,t_0)^\dag$ and
$\rho_S(t_0)=\rho_S(t_I)$ defined by Eq.(\ref{Eqn:ini_env}).
$\rho_{SE}(t_1)$ is in general not a product state with  fixed
environment state due to the system-environment interaction, this
means $\Lambda(t_2,t_1)$ is not a UDM in general. The history of the
system  from $t_0$ to $t_1$ is encoded in $\rho_{SE}(t_1)$.  We will
refer to this type of map as dynamical map with  memory. This
interpretation of the dynamical map is frequently used to
investigate the non-Markovinity of quantum dynamics.

In particular, when the starting time of the map is
the initial time of the evolution, i.e., $t_1=t_0$, $\Lambda(t_2,t_0)$
is a UDM represented by,
\begin{eqnarray}
\rho_S(t_2)&=&\Lambda(t_2,t_0)\rho_S(t_0)\nonumber
\\&=&\Tr_E[U(t_2,t_0)\rho_{S}(t_0)\otimes\rho_{E}(t_0)U(t_2,t_0)^{\dag}]
\label{Eqn:lambda0}
\end{eqnarray}
which can map any system state to another physical state.

\subsection{Type \rm{II}: memoryless dynamical map}
Next, we define $T(t_2,t_1)$ as the quantum dynamical map
transferring the quantum system from state $\rho(t_1)$ to
$\rho(t_2)$, where $t_1$ is the initial time of the evolution, i.e.,
$t_I=t_1$. The initial total state is
$\rho_{SE}(t_1)=\rho_S(t_1)\otimes\rho_{E}(t_1)$ with fixed
$\rho_E(t_1)$ defined by Eq.(\ref{Eqn:ini_env}) and
arbitrary $\rho_S(t_1)$. Then, $T(t_2,t_1)$ can be  expressed as,
\begin{eqnarray}
  \rho_S(t_2)&=&T(t_2,t_1)\rho_S(t_1)  \nonumber
\\&=&\Tr_E[U(t_2,t_1)\rho_S(t_1)\otimes\rho_E(t_1)U(t_2,t_1)^{\dag}].
\label{Eqn:T}
\end{eqnarray}
$T(t_2,t_1)$ is always a UDM in contrast to $\Lambda(t_2,t_1)$.
 As $\rho_E(t_1)=\rho_E(t_I)$ is independent of the system's history,
the map $T$ is memoryless. We remind the reader that in a quantum process,
$T(t_2,t_1)$ and $T(t'_2,t'_1)$ ($t_1\neq t'_1$) correspond to different evolutions.

In particular, when the total Hamiltonian $H$ is time-independent and
the initial environmental state is stationary, we
have
\begin{eqnarray}
\rho_S(t_2)&=&T(t_2,t_1)\rho_S(t_1) \nonumber
\\&=&\Tr_E[e^{-\frac{i}{\hbar}H(t_2-t_1)}\rho_S(t_1)
\otimes\rho_E e^{\frac{i}{\hbar}H(t_2-t_1)}]. \label{Eqn:homoT}
\end{eqnarray}
One can easily observe that the map $T(t_2,t_1)$ is time-homogeneous, i.e., $T(t_2,t_1)=T(t_2-t_1,0)$. This feature was
discussed by Chru\'{s}ci\'{n}ski and Kossakowski in Ref.
\cite{Chruscinski}, leading to the existence of an initial time $t_I$
in a non-Markovian time-local master equation,
\begin{eqnarray}
\dot{\rho}_S=\mathcal{L}(t-t_I)\rho_S,
\label{Eqn:MEwithTi}
\end{eqnarray}
such that
$T(t_2,t_1)=\mathcal{T}e^{\int_{t_1}^{t_2}\mathcal{L}(\tau-t_1)d\tau}=
\mathcal{T}e^{\int_{0}^{t_2-t_1}\mathcal{L}(\tau)d\tau}=T(t_2-t_1,0)$.
Eq.(\ref{Eqn:MEwithTi}) provide a full description of the quantum
non-Markovian process where evolutions starting at any time $t_I$
are included. The initial time $t_I$ is usually ignored in
literature provided that one only considers evolutions starting from
$t=0$,  leading to $\dot{\rho}=\mathcal{L}(t)\rho$ which may have the
same form of a time-dependent Markovian equation with non-negative
decay rates. However, the existence of the initial time $t_I$ is the
essential feature of a non-Markovian time-local equation compared
with a time-dependent Markovian equation. It provides an evident
signature of the memory effect since the dynamical map
$\varepsilon(t_2,t_1)=\mathcal{T}e^{\int_{t_1}^{t_2}\mathcal{L}(\tau-t_I)d\tau}$
sending $\rho_S(t_1)$ to $\rho_S(t_2)$ relies on the initial time
$t_I$ ($t_I\leq t_1\leq t_2$), i.e., the history of the
system, while in a time-dependent Markovian equation,
$\varepsilon(t_2,t_1)=\mathcal{T}e^{\int_{t_1}^{t_2}\mathcal{L}(\tau)d\tau}$
is uniquely defined.

When $t_1$ is the initial time of the evolution $t_0$,  we have
$\Lambda(t_2,t_0)=T(t_2,t_0)$ by definition. The key difference
between the two types of dynamical maps $\Lambda(t_2,t_1)$ and
$T(t_2,t_1)$ exists when $t_1$ is not the initial time.
When $t_1\neq t_0$, in general $\Lambda(t_2,t_1)\neq T(t_2,t_1)$
since $\rho_{SE}(t_1)$ is in general not $\rho_S(t_1)\otimes\rho_E(t_1)$
with fixed $\rho_E(t_1)$.
Given a quantum process characterized by the total Hamiltonian
and the initial environment state, the map $T$ is well-defined and
straightforward to calculate compared with $\Lambda$.
With these notations, the non-Markovianity can be
defined by the use of the memoryless dynamical map in the following
section.

\section{Non-Markovianity based on memory effects}

In this section, we give a condition as well as a  measure for
quantum non-Markovinity quantifying the degree of the memory effect.
Let us start from the quantum Markov process. A quantum process
is called Markovian if the corresponding dynamical map, which
we denote by $\Lambda_M$ in this paper, is divisible.
i.e.,
\begin{eqnarray}
 \Lambda_M(t+\tau)=\Lambda_M(\tau)\Lambda_M(t)
\label{Eqn:Mdiv}
\end{eqnarray}
where all maps are UDMs (CP and TP) defining a
one-parameter semigroups. Equivalently, the quantum Markovian process
can be described by a master equation in Lindblad form,
\cite{Lindblad}
\begin{eqnarray}
\dot{\rho}_S&=&\mathcal{L}\rho_S \nonumber\\
&=&-i[H,\rho_S]+\sum_\alpha\gamma_a(V_\alpha\rho_S {V_\alpha}^\dag
-\frac{1}{2}{V_\alpha}^\dag V_\alpha \rho_S\nonumber\\
&&-\frac{1}{2}\rho_S{V_\alpha}^\dag V_\alpha ), \label{Eqn:ME}
\end{eqnarray}
with $\gamma_{\alpha}\geq0$. The evolution of the system   is given
by $\rho_S(t_2)=\Lambda_M(t_2-t_1)\rho_S(t_1)$ where
$\Lambda_M(t)=e^{\mathcal{L}t}$. More generally, when the generator
$\mathcal{L}$ varies with time due to the change of external
conditions, the process is called time-dependent Markovian which
satisfies a time-dependent master equation in Lindblad form,
\begin{eqnarray}
\dot{\rho}_S&=&\mathcal{L}(t)\rho_S\nonumber\\
&=& -i[H(t),\rho_S]+\sum_\alpha\gamma_{\alpha}(t)
[V_\alpha(t)\rho_S{V_\alpha}^\dag(t)-\nonumber\\
&&\frac{1}{2}{V_\alpha}^\dag(t)V_\alpha(t)
\rho_S-\frac{1}{2}\rho_S{V_\alpha}^\dag(t) V_\alpha(t)]
\label{Eqn:TME}
\end{eqnarray}
with $\gamma_\alpha(t)\geq 0$. The dynamical map  becomes
$\Lambda_M(t_2,t_1)=\mathcal{T}e^{\int_{t_1}^{t_2}\mathcal{L}(\tau)d\tau}$
which is inhomogeneous but still a UDM, and the evolution of the
system can be described by
$\rho_S(t_2)=\Lambda_M(t_2,t_1)\rho_S(t_1)$. The divisibility
condition can then be written as
\begin{eqnarray}
  &&\Lambda_M(t_2,t_0)=\Lambda_M(t_2,t_1)\Lambda_M(t_1,t_0)
\label{Eqn:Tdiv}
\end{eqnarray}
where each dynamical map is a UDM.  We will discuss in the following the
time-dependent Markovian process without loss of generality.

In a quantum (time-dependent) Markovian process, the dynamical map
transferring $\rho_S(t_1)$ to $\rho_S(t_2)$ is uniquely given by the
UDM $\Lambda_M(t_2,t_1)$ regardless of the system's history before $t_1$.
We remark that the map
$\Lambda_M(t_2,t_1)$ is a UDM if and only if it is induced from a
product state, i.e.
\begin{eqnarray}
  \rho_S(t_2)&=&\Lambda_M(t_2,t_1)\rho_S(t_1) \nonumber
  \\&=&\Tr_E[U(t_2,t_1)\rho_{S}(t_1)\otimes\rho_{E}(t_1)U(t_2,t_1)^{\dag}]
\label{Eqn:TDMmap}
\end{eqnarray}
where $\rho_{E_I}(t_1)$ is a fixed state and $\rho_S(t_1)$ is
arbitrary. The evolution of the system at any time $t_1$ in
a  (time-dependent) Markovian process can be understood by examining
Eq.(\ref{Eqn:TDMmap}). It is interpreted as the collisional model as
reviewed in Ref.\cite{RivasR}. Particularly, when the total
Hamiltonian  and the initial environmental state are
time-independent, the map $\Lambda_M(t_2,t_1)$ can be understood as
\begin{eqnarray}
  \rho_S(t_2)&=&\Lambda_M(t_2,t_1)\rho_S(t_1) \nonumber
  \\&=&\Lambda_M(t_2-t_1)\rho_S(t_1)  \nonumber
  \\&=&\Tr_E[e^{-iH(t_2-t_1)}\rho_{S}(t_1)\otimes\rho_{E}e^{iH(t_2-t_1)}]
\label{Eqn:homoprd}
\end{eqnarray}
which depends only on the difference $t_2-t_1$, corresponding  to a
homogeneous Markovian process described by Eq.(\ref{Eqn:Mdiv}) or
Eq.(\ref{Eqn:ME}).

In a real Markovian quantum process, the total
state of the system and the environment may not remain perfectly factorized
as in Eq.(\ref{Eqn:TDMmap}), and $\varepsilon(t_2,t_1)$ is not a UDM
in general. From this point of view, the exact dynamics of an open
quantum system is never perfectly Markovian \cite{Rivas}. However,
if $\rho_{SE}(t)\approx\rho_S(t)\otimes\rho_E(t)$ with fixed
$\rho_E(t)$ (independent of the system state) where the correlation
between the system and environment does not affect so much the
system's dynamics, a Markovian model is still a good approximate
description.

An important observation is that the evolution from any  time $t_1$
in a quantum (time-dependent) Markovian process can be interpreted
as Eq.(\ref{Eqn:TDMmap}) where the total state $\rho_{SE}(t_1)$
remains a product state and the environmental state is fixed for arbitrary
$\rho_S(t_1)$. No matter the
initial time of an evolution is $t_1$ or $t_0$ (or any time before
$t_1$), the dynamical map transferring $\rho_S(t_1)$ to
$\rho_S(t_2)$ is always given by $\Lambda_M(t_2,t_1)$. Therefore,
the Markovian dynamical map $\Lambda_M$ can be interpreted as both
$\Lambda$ and $T$ according to our definitions in Sec.\rm{III},
i.e.,
\begin{eqnarray}
\Lambda_M(t_2,t_1)=\Lambda(t_2,t_1)=T(t_2,t_1)
\end{eqnarray}
for any $t_0\leq t_1\leq t_2$. Consequently, in a Markovian
process, the divisibility condition Eq.(\ref{Eqn:Tdiv}) can be
expressed both in terms of $\Lambda$ and $T$, respectively,
\begin{eqnarray}
 \Lambda(t_2,t_0)=\Lambda(t_2,t_1)\Lambda(t_1,t_0),\label{Eqn:NMdivl}\\
  T(t_2,t_0)=T(t_2,t_1)T(t_1,t_0).
  \label{Eqn:NMdivT}
\end{eqnarray}

Any violation of the above two divisibility conditions will be a
manifestation of non-Markovianity. Interestingly, the violations of
Eq.(\ref{Eqn:NMdivl}) and Eq.(\ref{Eqn:NMdivT}) have different
physical interpretations which can lead to different criteria and
measures for quantum non-Markovianity. The violation of
Eq. (\ref{Eqn:NMdivl}) is usually manifested  by the non-complete
positivity of the intermediate map $\Lambda(t_2,t_1)$ in evolutions
starting at $t_0$, which is the ultimate reason behind the behaviors
of different quantities (such as the trace distance,
entanglement, correlation and so on)
in Refs.\cite{Breuer,Laine,Vasile,Liuj,Rivas2,Lu,Luo,Lorenzo,
Bylicka,Rajagopal,Alipour}. The violation itself also accounts
for the measures in Refs. \cite{Rivas2,Hou,Hall,Chruscinski2014}.
In contrast, the dynamical maps in Eq.(\ref{Eqn:NMdivT})  are all
UDMs, i.e., CP and TP, and the violation of the second divisibility
condition Eq.(\ref{Eqn:NMdivT}) is manifested by the inequality
\begin{eqnarray}
&&T(t_2,t_0)\neq T(t_2,t_1)T(t_1,t_0). \label{Eqn:ineq}
\end{eqnarray}
A quantum process is non-Markovian if there  exists $t_2\geq t_1 \geq
t_0$ such that the inequality Eq.(\ref{Eqn:ineq}) holds. This
criterion is conceptually different from others. It has a clear
physical interpretation in terms of memory effects, which we will
explain in the following.

\begin{figure}
\includegraphics*[width=7.5cm]{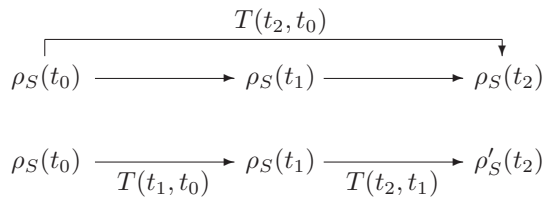}
\caption{Schematic illustration of the non-Markovnian inequality Eq.(\ref{Eqn:ineq}). }
\label{FIG:arrow}
\end{figure}

Consider the left-hand and right-hand sides of Eq.(\ref{Eqn:ineq})
acting on an arbitrary state $\rho_S(t_0)$, respectively, as shown
in Fig.\ref{FIG:arrow}. On the left-hand side of
Eq.(\ref{Eqn:ineq}), $\rho_S(t_0)$ is mapped  to $\rho_S(t_2)$ by
$T(t_2,t_0)$ in evolution A, while on the right-hand side,
$\rho_S(t_0)$ is firstly mapped to $\rho_S(t_1)$ by $T(t_1,t_0)$ in
evolution B. Then, as an initial state, $\rho_S(t_1)$ is mapped to
$\rho'_S(t_2)$ by $T(t_2,t_1)$ in evolution C, which starts at $t_1$
with $\rho_{SE}(t_1)=\rho_{S}(t_1)\otimes\rho_{E}(t_1)$. If
$T(t_2,t_0)\neq T(t_2,t_1)T(t_1,t_0)$, there exists $\rho_S(t_0)$
such that $\rho_S(t_2)\neq \rho'_S(t_2)$.  We remind the reader  that in both
evolutions A and C, the system state at $t_1$ is $\rho_S(t_1)$, but
with different histories: in evolution A, $\rho_S(t_1)$ has a
history from $t_0$ to $t_1$ which is taken into account by
$\rho_{SE}(t_1)=U(t_1,t_0)\rho_{S}(t_0)\otimes\rho_{E}(t_0)U(t_1,t_0)^{\dag}$.
Nevertheless, in evolution C, $\rho_S(t_1)$ serves as an initial
state without any history before $t_1$. Therefore, the fact that
$\rho_S(t_2)\neq \rho'_S(t_2)$ is a direct manifestation of the
memory effect where the future evolution (after $t_1$) of the system
depends on its history (from $t_0$ to $t_1$). This is the
fundamental property of non-Markovianity.  If $T(t_2,t_0)=
T(t_2,t_1)T(t_1,t_0)$ provided $t_2>t_1>t_0$, then
$\rho_S(t_2)=\rho'_S(t_2)$ for any $\rho_S(t_0)$, we say that the
future state of the system depends only on its present state, i.e.,
the process is Markovian.

From the   environment side, the physics  of Eq.(\ref{Eqn:ineq}) can
be interpreted as follows. At the end of evolution B, the total
state of the system and the environment is
$\rho_{SE}(t_1)=U(t_1,t_0)\rho_{S}(t_0)\otimes\rho_{E}(t_0)U(t_1,t_0)^{\dag}$.
Then, at the beginning of evolution C, the environment is
initialized by $T(t_2,t_1)$ such that
$\rho_{SE}(t_1)\rightarrow\rho_{S}(t_1)\otimes\rho_{E}(t_1)$, where
$\rho_E(t_1)$ is the initial environmental state at $t_1$ defined in
Eq.(\ref{Eqn:ini_env}). The terminology {\it initialize} means the
system information acquired by the environment in the time interval
$[t_0,t_1]$ is erased at time $t_1$. Note that the initialization
never happens in evolution A. After $t_1$, if the future system
states in evolutions A and B are different, that means the
environment "remembers" the history of the system [encoded in the
$\rho_{SE}(t_1)$] and the future system is affected by this kind of
memory.

We define the non-Markovianity as the maximum distance  between
$T(t_2,t_0)$ and $T(t_2,t_1)T(t_1,t_0)$ over all $t_1$ and $t_2$ (We
may investigate a quantum process with a fixed $t_0$),
\begin{eqnarray}
N_M=\max_{t_1,t_2} D(T(t_2,t_0), T(t_2,t_1)T(t_1,t_0))
\label{Eqn:NM}
\end{eqnarray}
Here $D$ denotes some distance measurement between  the two
quantum dynamical maps $T(t_2,t_1)$ and $T(t_2,t_1)T(t_1,0)$. (Here
the concatenation of the two UDMs $T(t_2,t_1)$ and $T(t_1,t_0)$ is
also a UDM). This definition allows us to quantify non-Markovianity
directly through dynamical maps without optimization of quantum
states \cite{Breuer}. It can be understood as the maximal deviation
of the divisibility condition Eq.(\ref{Eqn:NMdivT}). When
$N_M\rightarrow0$, a quantum process loses its memory effects and
becomes Markovian.

To choose a distance measure between two dynamical maps for our
non-Markovianity, we remark that a dynamical map $\Lambda$ is
isomorphic to its Choi-Jami\'{o}{\l}kowski matrix
\cite{Choi,Jamiolkowski} defined as $\rho_\Lambda=\Lambda\otimes
\mathbb{I}(\ket{\psi}\bra{\psi})$. Here $\mathbb{I}$ is the identity
map, and $\ket{\psi}=\frac{1}{\sqrt{n}}\sum_{i=1}^{n}\ket{i}\ket{i}$
is a maximally entangled state of the system and an ancillary
system. The Choi-Jami\'{o}{\l}kowski matrix has been used in
Ref.\cite{Rivas2} to quantify the non-complete positivity and
measure the non-Markovianity. Meanwhile, it is often used for
measuring the fidelity or distance between two quantum channels
(typically a general channel and a unitary one)
\cite{Gilchrist,Raginsky,WATROUS,Johnston}. Here we can easily
measure the distance between two dynamical maps $\Lambda_1$ and
$\Lambda_2$ through the trace distance of their
Choi-Jami\'{o}{\l}kowski matrices, i.e.,
\begin{eqnarray}
D(\Lambda_1,\Lambda_2)=\frac{1}{2}||\rho_{\Lambda_1}-\rho_{\Lambda_2}||
\label{Eqn:dis}
\end{eqnarray}
where $||A||=\Tr(\sqrt{A^\dag A})$ is the trace norm of  an operator
$A$. The good properties of the trace distance can be taken
advantage of for measuring the distance of dynamical maps
\cite{Gilchrist}. Finally, from Eq.(\ref{Eqn:NM}) and
Eq.(\ref{Eqn:dis}), the non-Markovianity is given by
\begin{eqnarray}
N_M=\max_{t_1,t_2}\frac{1}{2}||\rho_{T(t_2,t_0)}-\rho_{T(t_2,t_1)T(t_1,t_0)}||
\label{Eqn:NM2}
\end{eqnarray}
which naturally gives a finite value of non-Markovianity  satisfying
$0\leq N_M\leq1$ for any quantum process without normalization.

Given a theoretically described quantum process, the  dynamical map
$T(t_2,t_1)$ is always CP and describes a physically plausible
evolution. The determination of $T(t_2,t_1)$ (or its
Choi-Jami\'{o}{\l}kowski matrix) is straightforward as long as
the evolution starting from $t_1$ is known, regardless of
how the process is described. In experiment, $T(t_2,t_1)$
($\rho_{T(t_2,t_1)}$) could be determined through quantum process
tomography (quantum state tomography) \cite{Nielsen}. When the
complete information about $T(t_2,t_1)$ is unavailable,
$\rho_S(t_2)\neq \rho'_S(t_2)$ can be used as a sufficient condition
of non-Markovianity which is easy to verify both theoretically
and experimentally. Moreover, for any observable $A$,
$\Tr[A\rho(t_2)]\neq\Tr[A\rho'(t_2)]$ is sufficient for
non-Markovianity.


\section{Example}
In this section, we illustrate how our measure can be  calculated
with a typical quantum process. The model describes a two-level
system (denoted by S) decaying into its environment (denoted by E),
which is initially in the vacuum state. This model is exactly
solvable and extensively discussed to study the non-Markovian
behaviors. The total time-independent Hamiltonian is written as
\begin{eqnarray}
H=H_S+H_E+H_{SE}
\label{Eqn:Hmt}
\end{eqnarray}
where $H_S=\omega_0 \sigma^+\sigma^-$, $H_E=\sum_k\omega_k
b_k^{\dag}b_k$ are the free Hamiltonians of S and E,
and $H_{SE}=\sigma_{+}\sum_k g_k b_k+\sigma_{-}\sum_k g_k^* b_k^{\dag}$
denotes the coupling between the qubit and the environmental modes.
 In the case that
$\rho_{SE}(0)=\rho_S(0)\otimes\rho_E$ where $\rho_S(0)$ is arbitrary
and $\rho_E=\ket{0}_E\bra{0}_E$ is the vacuum state, the evolution
starting at $0$ in the interaction picture) can be
 expressed in terms of dynamical map $T$ as
\begin{eqnarray}
\rho_{S}(t)&=&T(t,0)\rho_{S}(0)\nonumber\\
&=&\left(
\begin{array}{cc}
|c(t)|^2\rho_{S11}(0) & c(t)\rho_{S12}(0)\\
c(t)^*\rho_{S21}(0) & 1-|c(t)|^2\rho_{S11}(0)\\
\end{array}
\right)
\label{Eqn:T0t}
\end{eqnarray}
where the function $c(t)$ satisfies
\begin{eqnarray}
\dot{c}(t)=-\int_0^{t}dt'f(t-t')c(t')
\label{Eqn:c_dif_int}
\end{eqnarray}
with the correlation function
$f(t-t_1)=\int d\omega J(w)e^{i(\omega_0-w)(t-t_1)}$\cite{Breuerbook} .

Since the total Hamiltonian is time-independent and the initial
environmental state $\rho_E$ is invariant under the total
Hamiltonian, the map $T$ is time homogeneous according to the
discussion in Sec.2, i.e., $T(t_2,t_1)=T(t_2-t_1,0)$. Thus, the
evolution starting at $t_1$
($\rho_{SE}(t_1)=\rho_S(t_1)\otimes\rho_E$) can be easily obtained
as
\begin{eqnarray}
\rho_{S}(t_2)=T(t_2,t_1)\rho_{S}(t_1)=T(t_2-t_1,0)\rho_S(t_1).
\label{Eqn:Tt1t2}
\end{eqnarray}
Alternatively, the evoltuion starting at $0$ can be  described by
the following master equation
\begin{eqnarray}
\dot\rho_{S}&=&\mathcal{L}(t)\rho_{S} \nonumber\\
&=&-\frac{i}{2}S(t)[\sigma^+\sigma^-,\rho_S]+
\gamma(t)(\sigma^-\rho_S\sigma^+-\frac{1}{2}\sigma^+\sigma^-\rho_S\nonumber\\
&&-\frac{1}{2}\rho_S\sigma^+\sigma^-)
\label{Eqn:ME0}
\end{eqnarray}
where $S(t)=-2\mathrm{Im}[\frac{\dot{c}(t)}{c(t)}]$ and
$\gamma(t)=-2\mathrm{Re}[\frac{\dot{c}(t)}{c(t)}]$.  According
to Eq.(\ref{Eqn:ME0}), $T(t,0)$ is given by
$T(t,0)=\mathcal{T}e^{\int_{0}^{t}\mathcal{L}(\tau)d\tau}$. We
stress here that Eq.(\ref{Eqn:ME0}) can only describe evolutions
starting at $0$ rather than an arbitrary time $t_I$.
Instead, the master equation describing the full quantum
process contains the initial time $t_I$ \cite{Chruscinski},
\begin{eqnarray}
\dot\rho_{S}&=&\mathcal{L}(t-t_I)\rho_{S}   \nonumber\\
&=&-\frac{i}{2}S(t-t_I)[\sigma^+\sigma^-,\rho_S]+
\gamma(t-t_I)(\sigma^-\rho_S\sigma^+ \nonumber\\
&&-\frac{1}{2}\sigma^+\sigma^-\rho_S -\frac{1}{2}\rho_S\sigma^+\sigma^-)
\label{Eqn:MEtI}
\end{eqnarray}
such that
$T(t_2,t_1)=\mathcal{T}e^{\int_{t_1}^{t_2}\mathcal{L}(\tau-t_1)d\tau}
=\mathcal{T}e^{\int_{0}^{t_2-t_1}\mathcal{L}(\tau)d\tau}$  which is
consistent with Eq.(\ref{Eqn:Tt1t2}).

Assume the spectral density of the bath are
Lorentzian and of the form $J_L(\omega)=\frac{1}{2\pi}
\frac{\gamma_0\lambda^2}{(\omega_0-\omega)^2+\lambda^2}$ where
$\lambda$ is connected to the environment correlation time $\tau_E$
by $\tau_E=1/\lambda$ and $\gamma_0$ determines the time scale of
the system by $\tau_S=1/\gamma_0$. The solution of
Eq.(\ref{Eqn:c_dif_int}) is given by $c(t)=e^{-\lambda t/2}
[\cosh(\frac{dt}{2})+\frac{\lambda}{d}\sinh(\frac{dt}{2})]$ with
$d=\sqrt{\lambda^2-2\gamma_0\lambda}$ \cite{Breuerbook}. Since
$c(t)$ is real, we have $S(t-t_I)=0$ in Eq.(\ref{Eqn:MEtI}).
Typically, the exact dynamics of an open quantum system  is not
Markovian as discussed above. In this example, when the environment
correlation time $\tau_E$ is much smaller than the system
characteristic time $\tau_S$, i.e., $\tau_E\ll\tau_S$, a Markovian
model can be a good approximation \cite{Breuerbook,Breuerrev}. Thus
the non-Markovianity of this quantum process can be characterized by
the parameter $R=\frac{\gamma_0}{\lambda}=\frac{\tau_E}{\tau_S}$.
When $R<\frac{1}{2}$, the process can be described by
Eq.(\ref{Eqn:MEtI}) with $\gamma(t-t_I)\geq0$. Interestingly, when
we only consider evolutions starting at 0, Eq.(\ref{Eqn:MEtI})
reduces to Eq.(\ref{Eqn:ME0}) which has the form of a time-dependent
Markovian equation Eq.(\ref{Eqn:TME}) ( but with different meanings
because $T(t_2,t_1)$ is undefined in Eq.(\ref{Eqn:ME0}) for
$t_1>0$). Therefore, the process with $R<\frac{1}{2}$ is called
Markovian by previously proposed measures that use evolutions
starting at a fixed time. It seems counter-intuitive that even
though $\tau_E$ is comparable to $\tau_S$, for example,
$R=\frac{\tau_E}{\tau_S}=0.49$, the process is still called
Markovian. Also, other intuitively non-Markovian models might be
called Markovian by previous measures \cite{Apollaro}. In contrast,
we will show that non-zero non-Markovianity (memory effects) always
exist in this exact model, even for $R<\frac{1}{2}$. In addition,
the non-Markovinianity tends to zero as $R\rightarrow0$. Thus our
measure reflects how valid the Born-Markov approximation is in this
model. 

\begin{figure}
\includegraphics*[width=8.5cm]{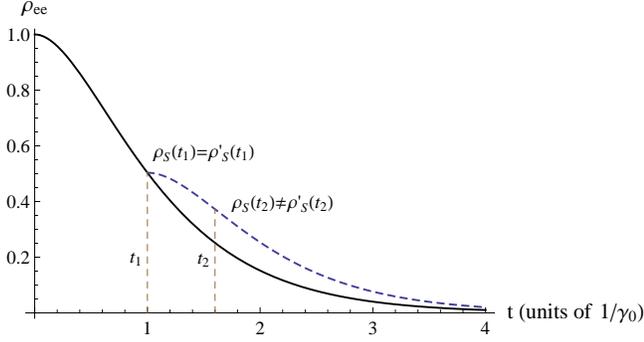}
\caption{(Color online) Evolutions of $\rho_{ee}$ with  different
starting time. The black solid line corresponds to the first
evolution starting at $0$, and the blue dashed line corresponds to
the second evolution starting at $t_1$. In the evolutions,
the system has the same state at $t_1$, however, they became
different at $t_2$, which is a manifestation of the memory effect. The
phenomenon never happens in a (time-dependent) Markovian process.}
\label{FIG:demon}
\end{figure}

We first demonstrate that when $R<0.5$, the evolution  in the
quantum process  depends on its history proving the process is
non-Markovian. Consider that the first evolution starts at $0$ where the
system is initially in its excited state, i.e.,
$\rho_{SE}(0)=\ket{e}\bra{e}\otimes\rho_E$. Then, the system density
matrix at $t_1$ is given by $\rho_S(t_1)=T(t_1,0)\rho_S(0)$. Now we
assume the second evolution starts at $t_1$ with initial state
$\rho'_S(t_1)=\rho_S(t_1)$ such that
$\rho'_{SE}(t_1)=\rho'_S(t_1)\otimes\rho_E=\rho_S(t_1)\otimes\rho_E$.
At a further time $t_2$, we have $\rho_S(t_2)=T(t_2,0)\rho_S(0)$ in
the first evolution and $\rho'_S(t_2)=T(t_2,t_1)\rho'_S(t_1)$ in the
second evolution. The result is visualized in Fig.\ref{FIG:demon} by
evaluating the evolutions of the excited-state population
$\rho_{ee}$ with $R=0.4$.  From the fact that
$\rho'_S(t_2)\neq\rho_S(t_2)$, we conclude that the future states (after
$t_1$) of the system are relevant to its history (from $t_0$ to
$t_1$) in the process. The exited-state population $\rho_{ee}$ decays
monotonically and non-exponentially in this case. Although
the decay rate $\gamma(t-t_I)$ is non-negative and the revival of
$\rho_{ee}$ does not occur, the environment is affected by the
system's history and then has an influence on the future evolution
of the system. Thus the dashed line and the solid line in
Fig.\ref{FIG:demon} do not overlap. This phenomenon never happens in
a (time-dependent) Markovian process described by Eq. (\ref{Eqn:TME}),
where the dynamical map transferring $\rho_S(t_1)$ to $\rho_S(t_2)$
is uniquely given by
$\Lambda_M(t_2,t_1)=\mathcal{T}e^{\int_{t_1}^{t_2}\mathcal{L}(\tau)d\tau}$
regardless of the initial time.

\begin{figure}
\includegraphics*[width=7.5cm]{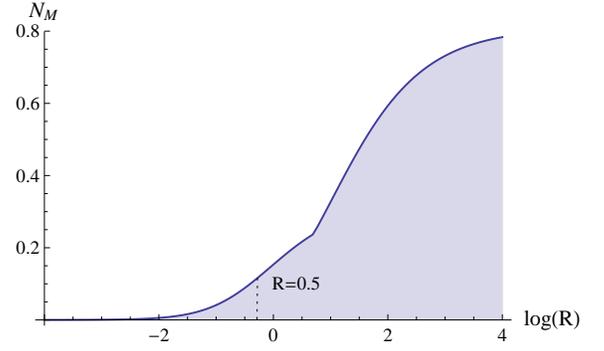}
\caption{(Color online) Non-Markovianity as a function of $\log{R}$
corresponding to the Lorentzian spectral density. It is shown that
$N_M$ is always non-zero for $R>0$ by our measurement.}
\label{FIG:NMvsR}
\end{figure}

The non-Markovianity $N_M$ for this quantum process is  calculated
as follows. From Eq.(\ref{Eqn:T0t}) and Eq.(\ref{Eqn:Tt1t2}), we
obtain the Choi-Jami\'{o}{\l}kowski matrix of $T(t_2,t_0)$
\begin{eqnarray}
\rho_{T(t_2,t_0)}&=&T(t_2,t_0)\otimes\mathbb{I}(\ket{\psi}\bra{\psi})\\
&=&\frac{1}{2}\left(
\begin{array}{cccc}
|c(\tau_{20})|^2 & 0 & 0 & c(\tau_{20})\\
0 & 0 & 0 & 0 \\
0 & 0 & 1-|c(\tau_{20})|^2 & 0 \\
c(\tau_{20})^* & 0 & 0 & 1 \\
\end{array}
\right)\nonumber
\label{Eqn:CM12}
\end{eqnarray}
where $\tau_{20}=t_2-t_0$. Similarly, the  Choi-Jami\'{o}{\l}kowski
matrix of $T(t_2,t_1)T(t_1,t_0)$ is
\begin{eqnarray}
&&\rho_{T(t_2,t_1)T(t_1,t_0)}\\
&=&T(t_2,t_1)T(t_1,t_0)\otimes\mathbb{I}(\ket{\psi}\bra{\psi})\nonumber\\
&=&\frac{1}{2}\left(
\begin{array}{cccc}
|c(\tau_{21})|^2|c(\tau_{10})|^2 & 0 & 0 &
\!\!\!\!\!\!\!\!\!\!\! \!\!\!\!\!\!\!\!\!\!
c(\tau_{21})c(\tau_{10})\\
0 & 0 & 0 & 0 \\
0 & 0 & 1\!\!-\!\!|c(\tau_{21})|^2 |c(\tau_{10})|^2 & 0 \\
c(\tau_{21})^* c(\tau_{10})^* & 0 & 0 & 1 \\
\end{array}
\right)\nonumber
\label{Eqn:CM0112}
\end{eqnarray}
where $\tau_{21}=t_2-t_1$ and $\tau_{10}=t_1-t_0$.  According to
Eq.(\ref{Eqn:NM2}), the non-Markovnianity in the case of the
Lorentz spectrum is
\begin{eqnarray}
N_M&=&\max_{\tau_{10},\tau_{21}}[\ \frac{1}{8}|M(N+\sqrt{4+N^2})|+\nonumber\\
&&\frac{1}{8}|M(N-\sqrt{4+N^2})|+\frac{1}{4}|MN|\ ]
\label{Eqn:NMans}
\end{eqnarray}
where $M=c(\tau_{20})-c(\tau_{10})c(\tau_{21})$,
$N=c(\tau_{20})+c(\tau_{10})c(\tau_{21})$ and
$\tau_{20}=\tau_{21}+\tau_{10}$. Notice that this result has been
simplified by the fact that $c(t)$ is real.

We calculate the non-Markovianity by numerically  optimizing the two
time differences $\tau_{10}$ and $\tau_{21}$. $N_M$ is plotted as a
function of $\log{R}$ in Fig.\ref{FIG:NMvsR} where $R$ varies from
$10^{-4}$ to $10^{4}$. The result demonstrates that the
non-Markoviaity is non-zero for all $R>0$. It monotonically
decreases with $R$ and tends to $0$ as $R\rightarrow0$. It is
observed that when $R<10^{-2}$, the non-Markovianity is already very
small ($N_M<0.006$), implying that the quantum process is approaching
Markovian and almost memoryless. Indeed, when $N_M$ is small, the
dashed and solid lines in Fig.\ref{FIG:demon} will be very close and
almost exponential. Then, a Markovian master equation can well
describe evolutions starting from any time,  i.e., the full
quantum process. In this model, the non-exponential relaxation is a
sign of non-Markovianity.

Before closing this section,  we consider the Ohmic spectral
density with an exponential cutoff $J_{O}(\omega)=\alpha w
 e^{-\frac{\omega}{\omega_c}}$, where $\alpha$ is a dimensionless
coupling strength and $\omega_c$ is the cutoff frequency. In this
case, the full dynamics is still described by Eqs.(\ref{Eqn:Tt1t2})
or (\ref{Eqn:MEtI}), whereas the function $c(t-t_I)$ is complex  and
$S(t-t_I)$ is non-zero in general. For complex $c(t)$, the
non-Markovianity has the following form,
\begin{eqnarray}
N_M&=&\max_{\tau_{10},\tau_{21}}[\ \frac{1}{8}|M'-\sqrt{M'^2+4(N'-2K)}|+\nonumber\\
&&\frac{1}{8}|M'+\sqrt{M'^2+4(N'-2K)}|+\frac{1}{4}|M'|\ ]
\label{Eqn:NMcplx}
\end{eqnarray}
where $M'=|c(\tau_{20})|^2-|c(\tau_{10})c(\tau_{21})|^2$,
$N'=|c(\tau_{20})|^2+|c(\tau_{10})c(\tau_{21})|^2$,
and $K=\mathrm{Re}[c(\tau_{20})c^*(\tau_{10})c^*(\tau_{21})]$.

\begin{figure}
\includegraphics*[width=7.5cm]{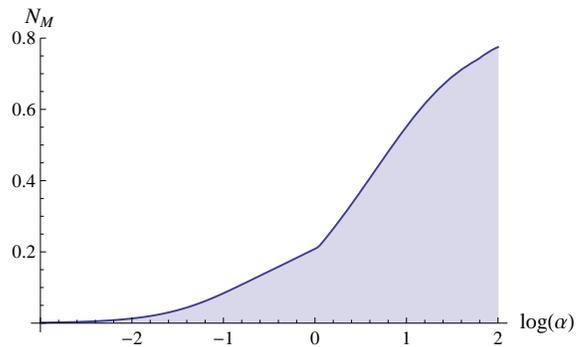}
\caption{(Color online) Non-Markovianity as a function of $\log{\alpha}$
corresponding to the Ohmic spectral density with $\omega_0/\omega_c=1$.
Similar to the result in FIG.(\ref{FIG:NMvsR}), $N_M$ is always non-zero for
 $\alpha>0$.}
\label{FIG:NMalpha}
\end{figure}

The exact value of $c(t-t_I)$ for the Ohmic spectrum
can be calculated using the analytic expressions in Ref.\cite{Zhang},
that is,
\begin{eqnarray}
c(t)&=&e^{i\omega_0t}\{\mathcal{Z}e^{-i\omega't}+  \nonumber \\
&&\frac{2}{\pi}\int_0^{\infty}d\omega\frac{J_O(\omega)e^{-i\omega t}}
{4[\omega-\omega_0-\Sigma(\omega)]^2+J_O^2(\omega)}\}
\label{Eqn:OhmicC}
\end{eqnarray}
where $\Sigma(\omega)=\alpha\omega_c[\frac{\omega}{\omega_c} e^{-
\frac{\omega}{\omega_c}}\mathrm{Ei}(\frac{\omega}{\omega_c})-1]$.
When the condition $\alpha\omega_c>w_0$ holds,
$\mathcal{Z}=[1-\Sigma'(\omega')]^{-1}\neq0$ corresponding to a
dissipationless process ($c(\infty)\neq0$) due to the zero spectral
density for negative frequencies, otherwise, $\mathcal{Z}=0$. Here
$w'$ is the solution of $\omega'=w_0-\Sigma(\omega')<0$. The global
phase $e^{iw_0t}$ in Eq.(\ref{Eqn:OhmicC}) is added compared with
the expression in Ref.\cite{Zhang} since we are working in the
interaction picture for consistency with the Lorentzian case. In
fact, the non-Markovianity does not depend on the picture we choose.
The non-Markovianity for different $\alpha$ (from $10^{-3}$ to
$10^2$)
 is calculated with $\omega_0/\omega_c=1$, as shown in
Fig.\ref{FIG:NMalpha}. Although the system dynamics for the Ohmic
spectrum is different from that for the lorentzian one, especially
in the strong-coupling regime, the behavior of $N_M$ as a function
of the coupling strength is similar. The non-Markovianity (memory
effect) is non-zero for all $\alpha>0$ even for parameters leading
to $\gamma(t-t_I)>0$ in Eq.(\ref{Eqn:MEtI}). When the coupling is
weak,  $\log(N_M)$ and   $\log{R}$ (or $\log{\alpha}$) are in a linear
relationship  in both cases, which indicates  that the quantum
process asymptotically becomes Markovian with the decreasing of the
coupling strength.

\section{Conclusions and Discussions}
We present  a criterion and propose  a measure for non-Markovianity
of quantum processes. The measure directly quantifies the degree of
memory effects, i.e., how much the future state of a system depends
on its past. To construct the measure, we introduce a universal
dynamical map (UDM) $T(t_2,t_1)$ which corresponds to an evolution
starting from $t_1$. In contrast to $\Lambda(t_2,t_1)$, $T(t_2,t_1)$
is well defined and  always CP as well as easy to calculate in a
quantum process. The Markovian divisibility can be expressed in
terms of $T$ in Eq.(\ref{Eqn:NMdivT}) and the violation of it is
simply manifested by the inequality Eq.(\ref{Eqn:ineq}), which has a
clear physical interpretation as the memory effects. We define
non-Markovianity as the maximal violation of Eq.(\ref{Eqn:NMdivT})
for all times. Unlike the previous proposed measures which focus on
evolutions starting from a fixed time, our measure applies to a
quantum process where evolutions can starting at an arbitrary time.

One important result of our work is that a quantum  process may have a
memory effect even if it is called Markovian by other measures.
Thus, the previously proposed criteria is not equivalent to the
memory effect. Besides, we   demonstrate that in a non-Markovian
process, the dynamics may be still described by a time-dependent
master equation in Lindblad-like form with non-negative decay rates.
However, the non-Markovian master equation contains the initial time
$t_I$, which is essentially different from a time-dependent Markovian
equation \cite{Chruscinski}. When only describing evolutions
starting from $0$ ($t_I=0$), a non-Markovian master equation may
have the same form of a time-dependent Markovian master equation,
e.g., Eq.(\ref{Eqn:ME0}) with $R<\frac{1}{2}$. By observing
evolutions starting from different times as in Fig.\ref{FIG:demon},
memory effects can be revealed. Thus the negative decay rates in a
master equation are not necessary to describe  memory effects
(non-Markovianity). And non-exponential but monotonic relaxation may
occur in both non-Markovian processes and time-dependent Markovian
processes.

Our measure is in units of trace distance that satisfies  $0\leq
N_M\leq 1$ for any quantum process without normalization.
It is easy to calculate regardless of the description of the quantum
process. The optimization for quantum states or the knowledge of the
environmental state is not required. When
 the full information of the dynamical map $T$ is
unavailable, the condition $\rho_S(t_2)\neq \rho'_S(t_2)$
[$\Tr[A\rho(t_2)]\neq\Tr[A\rho'(t_2)]$] can be used as an witness of
non-Markovianity, which is easy to be examined both theoretically
and experimentally.

Now we consider the renormalized spectral density with
zero negative frequency components in the example $J'_L(\omega)=J_L(\omega)\theta(\omega)$ where $\theta(\omega)$ is the step function. By numerical simulation, we find that the dynamics and the non-Markovianity for $J_L(\omega)$ are different from those for $J'_L(\omega)$.
The influence caused by a negative component strongly depends on $\frac{\lambda}{\omega_0}$. When $\frac{\lambda}{\omega_0}$ are small, $J'_L(\omega)$ and $J_L(\omega)$ leads to almost the same dynamics and non-Markovianity for both strong coupling (large $R$) and weak coupling. When $\frac{\lambda}{\omega_0}$ are
large, the negative frequency of the Lorentzian spectrum alters the dynamics and the non-Markovianity significantly for all coupling strengths (even for weak couplings). The reason is that $J_L(\omega)$ is a symmetric function with $x=\omega_0$ the axis of symmetry and $\lambda$ the peak width. Thus, $\frac{\lambda}{\omega_0}$ determines the weight of the component of the negative frequency which in turn alters the correlation function and the dynamics evidently, whereas the non-Markovinity for $J'_L(\omega)$ is still non-zero for all $R>0$ and tends to $0$ as $R\rightarrow0$.

\section{ACKNOWLEDGMENTS}
We thank J. Cheng, H. Z. Shen and J. Nie for helpful
discussions. This work is supported by the National Natural
Science Foundation of China under Grant No. 11175032 and
No. 61475033.

\end{document}